\documentclass[9pt,twocolumn,letterpaper]{article}
\usepackage{spconf,amsmath,graphicx}
\usepackage{enumitem}
\usepackage{graphicx}  
\usepackage{amsmath}
\usepackage{tikz}
\usepackage{cite}
\usepackage{subfigure}
\usepackage{booktabs}
\usepackage{multirow}
\usepackage{amsmath}
\usepackage{graphicx}
\usepackage{array}
\usepackage{multirow}
\usepackage{booktabs}
\usepackage{colortbl}
\usepackage{xcolor}
\usetikzlibrary{arrows.meta, positioning}
\usepackage{url}

\title{Deep joint source-channel coding for wireless point cloud transmission}
\name{Cixiao Zhang$^{\star}$, Mufan Liu$^{\star}$, Wenjie Huang, Yin Xu, Yiling Xu, Dazhi He \thanks{$^{\star}$Equal contribution. Corresponding author: Yin Xu and Yiling Xu.}}
\address{\small Cooperative Medianet Innovation Center, Shanghai Jiao Tong University, China\\ \small Email: xuyin@sjtu.edu.cn, yl.xu@sjtu.edu.cn}
\begin{document}
%
\maketitle
\begin{abstract}
\small The growing demand for high-quality point cloud transmission over wireless networks presents significant challenges, primarily due to the large data sizes and the need for efficient encoding techniques. In response to these challenges, we introduce a novel system named \textit{Deep Point Cloud Semantic Transmission (PCST)}, designed for end-to-end wireless point cloud transmission. Our approach employs a progressive resampling framework using sparse convolution to project point cloud data into a semantic latent space. These semantic features are subsequently encoded through a deep joint source-channel (JSCC) encoder, generating the channel-input sequence. To enhance transmission efficiency, we use an adaptive entropy-based approach to assess the importance of each semantic feature, allowing transmission lengths to vary according to their predicted entropy. PCST is robust across diverse Signal-to-Noise Ratio (SNR) levels and supports an adjustable rate-distortion (RD) trade-off, ensuring flexible and efficient transmission. Experimental results indicate that PCST significantly outperforms traditional separate source-channel coding (SSCC) schemes, delivering superior reconstruction quality while achieving over a 50\% reduction in bandwidth usage.
\end{abstract}

\section{Introduction}

\begin{figure*}[t]
    \centering
    \includegraphics[width = 0.90\linewidth, height=0.43\linewidth]{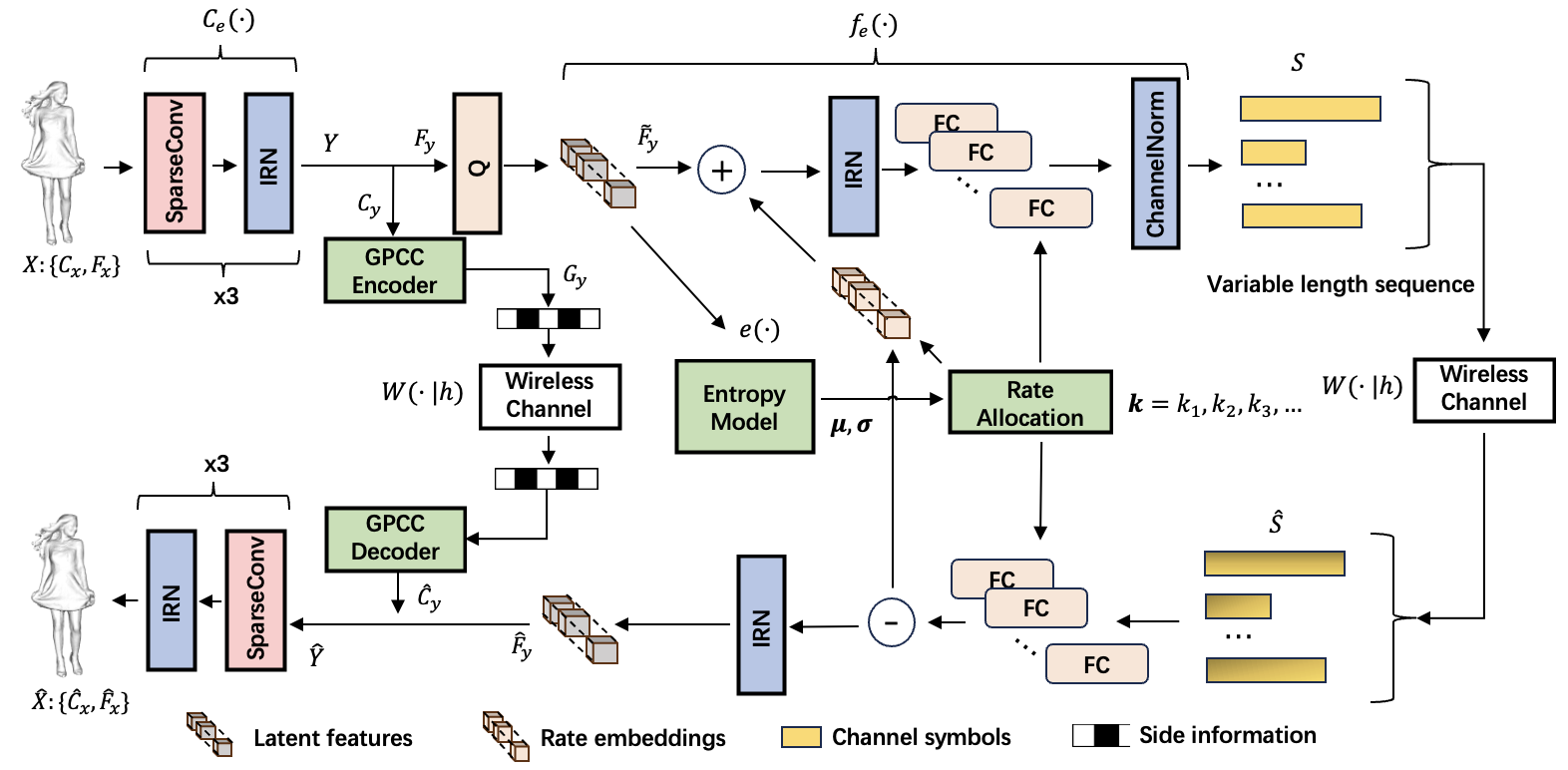}
    \caption{\small The PCST encoder has two main modules: the \textit{progressive multiscale resampling network} $C_e$ for extracting latent features, and the \textit{switchable fully connected layer} $f_e$ for encoding these features into variable-length symbols. The decoder mirrors the encoder’s structure.}
    \label{fig:enter-label}
    \vspace{-0.25cm}
\end{figure*}
\small Volumetric media, including point clouds and meshes, has gained significant traction recently, offering immersive six degree-of-freedom (6-DoF) experiences \cite{LDPC_ATSC}. This advancement aligns with the emergence of next-generation mobile networks, enhancing user experiences in the metaverse. However, transmitting high-quality point clouds demands exceptionally high throughput and reliability, often requiring several Gbps, which exceeds the capabilities of current 5G networks. Traditionally, point cloud transmission utilizes separate source-channel coding (SSCC). Source coding techniques, such as geometry-based point cloud compression (GPCC) and video-based point cloud compression (VPCC), aim to minimize redundancy by using methods like octree, trisoup, or converting 3D data into 2D video formats. Following compression, channel coding (e.g., LDPC \cite{LDPC_ATSC}, Polar code \cite{polar}) adds redundancy to protect against transmission errors. However, SSCC’s effectiveness is contingent on long code blocks, leading to considerable transmission delays, which are detrimental to real-time data transmission. Additionally, cliff effects arise when the channel coding performance collapses if the capacity is insufficient for the required communication rate.
\par JSCC (Joint Source-Channel Coding) has gained attention due to the limitations of SSCC. The autoencoder-based framework has been successful in transmitting semantic information across various media, including images, videos, and audio, thanks to its robust data reconstruction capabilities \cite{vae, JSCC1, JSCC, speech, dvst, deeptext}. Additionally, JSCC improves transmission efficiency by integrating entropy models that assess the significance of source data \cite{PCGCv2, Yan2019}. However, our review indicates a gap in the application of semantic transmission frameworks for volumetric media like point clouds. This gap exists because point clouds are characterized by large data volumes and irregular, sparse structures in 3D space, which pose challenges for traditional media processing frameworks. Additionally, the large volume of point clouds reduces neural network inference speed, impacting real-time rendering and immersive communication experiences.
\par To fill this research gap, we introduce the first \textit{Deep Point Cloud Semantic Transmission (PCST)} framework, leveraging the point cloud compression structure from \cite{PCGCv2}. Specifically, we utilize a progressive resampling framework based on sparse convolution to extract semantic features from point clouds. We then incorporate a switchable fully-connected (FC) layer module for variable-length encoding of these semantic features. This process is guided by an entropy model to prioritize features with varying levels of importance. Through the joint optimization of source and channel codes, PCST aims to achieve a compact representation of point clouds while maintaining robustness under different channel conditions. Our current study primarily focuses on the transmission of static point cloud geometry (PCG). Future work will explore the transmission of dynamic and attribute-attached point clouds, expanding the scope of our research. Extensive experimental results show that PCST consistently outperforms all SSCC schemes regarding reconstruction quality and bandwidth usage under both AWGN and Rayleigh channels.

\section{The proposed method}

\small 
\subsection{Architecture}
PCST consists of several components: a progressive multiscale resampling network (MultiRes), a deep JSCC codec, a GPCC codec, and an entropy model. The point cloud geometry (PCG) $X$ is represented as a sparse tensor $\{C_x, F_x\}$, where $C_x$ contains the occupied coordinates and $F_x$ contains feature attributes. Here, $F_x$ is an all-ones vector to indicate occupancy. In the PCST encoder, $X$ is first processed through the MultiRes encoder $c_e$, which performs progressive down-sampling to extract the latent features, resulting in a sparse tensor $Y = \{C_y, F_y\}$. The attributes $F_y$ of the latent features are quantized into $\Tilde{F}_Y$ and fed into both the entropy model and the Deep JSCC encoder $f_e$. The coordinates $C_y$ are compressed via the GPCC encoder $h_e$ and transmitted as side information. The entropy model estimates the entropy of each latent feature and assigns a code length proportional to its rounded entropy value. This guides the Deep JSCC encoder $f_e$ to produce a variable-length symbol sequence $S$. In our work, we evaluate the transmission overhead using the channel bandwidth ratio (CBR). It is defined as the ratio of the total length of all transmitted symbols, including both encoded latent features $S$ and side information, to the total dimensionality of the original point cloud (product of the number of points and the coordinate dimensions). All the symbols are transmitted over the wireless channel, e.g., $\hat{S} = W(S|\mathbf{h}) = \mathbf{h}\odot S+\mathbf{n}$, where $\mathbf{h}$ represents the channel state information (CSI) and $\mathbf{n}$ is the noise vector. The decoding process mirrors the encoding architecture, starting with $f_d$ and the GPCC decoder to reconstruct the sparse tensor $Y$ of latent features. The MultiRes decoder $c_d$ then performs progressive upsampling to restore the original PCG $\hat{X}$. The total data flow can be summarized as follows:
 \begin{equation}
\begin{minipage}{1\linewidth}
\centering
\vspace{-0.3cm}
\begin{tikzpicture}[node distance=0.8cm, auto, >=Latex]
    \node (X) {$X$};
    \node (Y) [right=of X] {$Y$};
    \node (Ytilde) [right=of Y] {$\Tilde{F}_y$};
    \node (S) [right=of Ytilde] {$S$};
    \node (Shat) [right=of S] {$\hat{S}$};
    \node (Yhat) [right=of Shat] {$\hat{F}_y$};
    \node (Xhat) [right=of Yhat] {$\hat{X}$.};
    \node (CY) [below=of Y] {$C_y$};
    \node (gpc) [right =of CY] {$G_y$};
    \node (gpd) [right =of gpc] {$\hat{G_y}$};
    \node (CY2) [right =of gpd] {$\hat{C_y}$};

    \draw[->] (X) -- node[above] {$c_e(\cdot)$} (Y);
    \draw[->] (Y) -- node[above] {$q(\cdot)$} (Ytilde);
    \draw[->] (Ytilde) -- node[above] {$f_e(\cdot)$} (S);
    \draw[->] (S) -- node[above] {$W(\cdot|h)$} (Shat);
    \draw[->] (Shat) -- node[above] {$f_d(\cdot)$} (Yhat);
    \draw[->] (Yhat) -- node[above] {$c_d(\cdot)$} (Xhat);
    
    \draw[->] (gpc) -- node[above] {$W(\cdot|h)$} (gpd);
    \draw[->] (gpd) -- node[above] {$h_d(\cdot)$} (CY2);

    \draw[->, bend left=45] (Y) -- (CY);
    \draw[->] (CY) -- node[above] {$h_e(\cdot)$} (gpc);
    \draw[->, bend right=45] (CY2) to (Yhat);
\end{tikzpicture}
\end{minipage}
\end{equation}
\vspace{-0.6cm}
\begin{equation}
\begin{minipage}{\linewidth}
\centering
\begin{tikzpicture}[node distance=0.8cm, auto, >=Latex]
    \node (Y2) {$F_y$};
    \node (MuSigma) [right=of Y2] {$\{\boldsymbol{\mu}, \boldsymbol{\sigma}\}$};
    \node (k) [right=of MuSigma] {$\boldsymbol{k}$.};

    \draw[->] (Y2) -- node[above] {$e(\cdot)$} (MuSigma);
    \draw[->] (MuSigma) -- node[above] {$g(\cdot)$} (k);
\end{tikzpicture}
\end{minipage}
\end{equation}
As side information is crucial for PCG reconstruction, it should be transmitted losslessly in wireless channel.


\subsection{Progressive Multiscale Resampling Network}
\vspace{-0.2cm}
\small Point cloud data is characterized by its large volume and rich informational content. Typically, acquired point cloud data is extremely sparse and unordered, making it unsuitable for conventional image processing techniques. In recent years, sparse convolution has proven to be highly efficient and well-suited for point cloud tasks due to its inherent sparsity \cite{pointnet}. This method has demonstrated excellent performance in tasks such as segmentation and recognition. Inspired by its effectiveness, we employ a \textit{multiscale progressive resampling} framework utilizing \textit{sparse convolution} \cite{Sparse} to extract deep semantic features (a.k.a., latent features) from PCG. This hierarchical process comprises three stages of downsampling and upsampling. Before feeding the point cloud into the PCST network, voxelization is performed to ensure that only voxels are used during convolution. Each downsampling step reduces the scale of each geometric dimension by half using sparse convolution with a stride of two. To enhance feature extraction efficiency, Inception-Residual Networks (IRN) \cite{irn} are integrated into each downsampling step. Following progressive downsampling, the coordinates and attributes of the latent features are obtained for wireless transmission. The coordinates of the latent features are compressed with a GPCC encoder, consuming only a small fraction of the total bitstream, while the attributes of the latent features are encoded using our deep JSCC encoder. For decoding, transposed sparse convolution with a stride of two is employed for each upsampling step. Similarly, IRN is incorporated for feature refinement. During each upsampling step, we reconstruct the geometric details by removing incorrect voxels and identifying accurately filled voxels through binary classification, determining whether each voxel has been occupied by a point. To establish the decision threshold $p$ for binary classification, or top-$k$ generation, we transmit the total number $N$ of reconstructed points as side information and divide it by the number of voxels.

\begin{figure*}[!t]
    \centering
    \subfigure[Ground truth]{
        \includegraphics[width=0.234\textwidth]{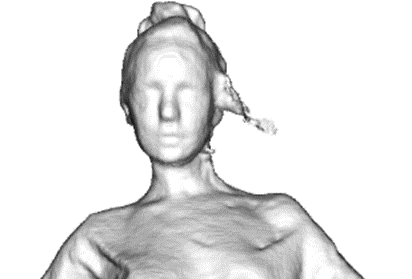}
        \label{Ground Truth}
    }
    \subfigure[\parbox{0.9in}{\centering Octree+LDPC\\CBR=0.060\\ PSNR=67.98dB}]{
        \includegraphics[width=0.235\textwidth]{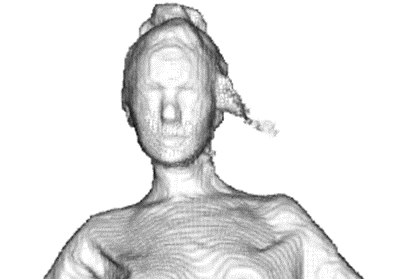}
        \label{fig:subfig2}
    }
    \subfigure[\parbox{0.9in}{\centering Trisoup+LDPC\\CBR=0.032\\PSNR=71.63dB}]{
        \includegraphics[width=0.235\textwidth]{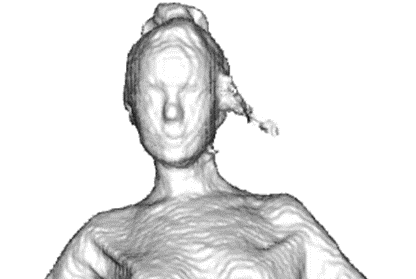}
        \label{fig:subfig3}
    }
    \subfigure[\parbox{0.9in}{\centering PCST\\CBR=0.027\\PSNR=73.60dB}]{
        \includegraphics[width=0.235\textwidth]{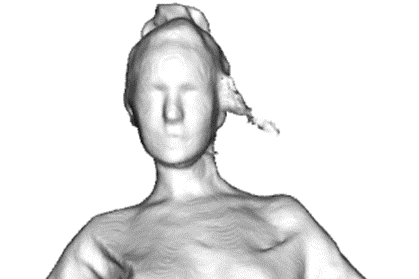}
        \label{fig:subfig4}
    }
    \vspace{-0.3cm}
    \caption{Visualization of PCG reconstruction.}
    \label{vis}
    \vspace{-0.6cm}
\end{figure*}

\subsection{Entropy-guided Deep JSCC}
\small \;\;\;\; \textbf{Structure:} Our deep JSCC encoder $f_e$ comprises an IRN and a switchable FC layer module, as illustrated in Fig. \ref{fig:enter-label}. IRN is utilized to extract JSCC-related semantic features before feeding them into the FC layers. We also tried with other neural networks, such as transformers and graph neural networks (GNNs), but found them not efficient as IRN for feature extraction. To allow JSCC-aware feature extraction, each latent feature is concatenated with the corresponding rating embedding before feeding into the IRN, as illustrated in Fig. \ref{fig:enter-label}. The switchable FC layer module is constructed to maps the latent features into channel symbols of variable lengths. It consists of a list of FC layers with identical input sizes but varying output sizes. Each FC layer's output length corresponds to one possible entropy of the latent features. The decoder of deep JSCC mirrors the encoder structure, reconverting the noisy channel symbols back to the latent features.

\par\textbf{Entropy model:} After extracting the latent features of the PCG, we follow the structure in \cite{dvst} to allocate bandwidth for latent feature transmission. We use the entropy model from \cite{hyper} to predict the entropy of the latent feature attributes $F_y$ and subsequently determine the necessary symbol length for deep JSCC encoding. This helps accommodate the semantic importance of various latent features in relation to channel coding. Specifically, we consider the latent feature attributes $F_y$ as a Gaussian Mixture Model (GMM) $\mathcal{N}(\mu_i, \sigma_i^2)$. The distribution is parameterized by a fully factorized density as
\begin{equation}
    P_{\psi}(F_y) = \prod_{i} \left(P_{\psi} * \mathcal{U}\left(-\frac{1}{2}, \frac{1}{2}\right) \right)(F_{Y_i}),
\end{equation}
where $*$ is the convolution operation, $P_{\psi}(F_y)$ represents the probability distribution of $F_y$ parameterized by $\psi$, and $\mathcal{U}\left(-\frac{1}{2}, \frac{1}{2}\right) $ denotes the uniform distribution. If the entropy model predicts that $F_{Y_i}$ has high entropy, the encoder will add more channel bandwidth for its transmission. The channel bandwidth cost $K$ for transmitting semantic features is calculated as
\begin{equation}
    K=\sum_i Q(k_i)=\sum_i Q(-\eta_i\log_2 P(F_{Y_i})),
\end{equation}
where $\eta_i$ is the controlling factor to trade-off channel bandwidth cost of latent feature attributes $F_{Y_i}$. $Q$ denotes the scalar quantization. 

\begin{figure*}[t]
    \centering
    \subfigure[]{
        \includegraphics[width=0.235\textwidth]{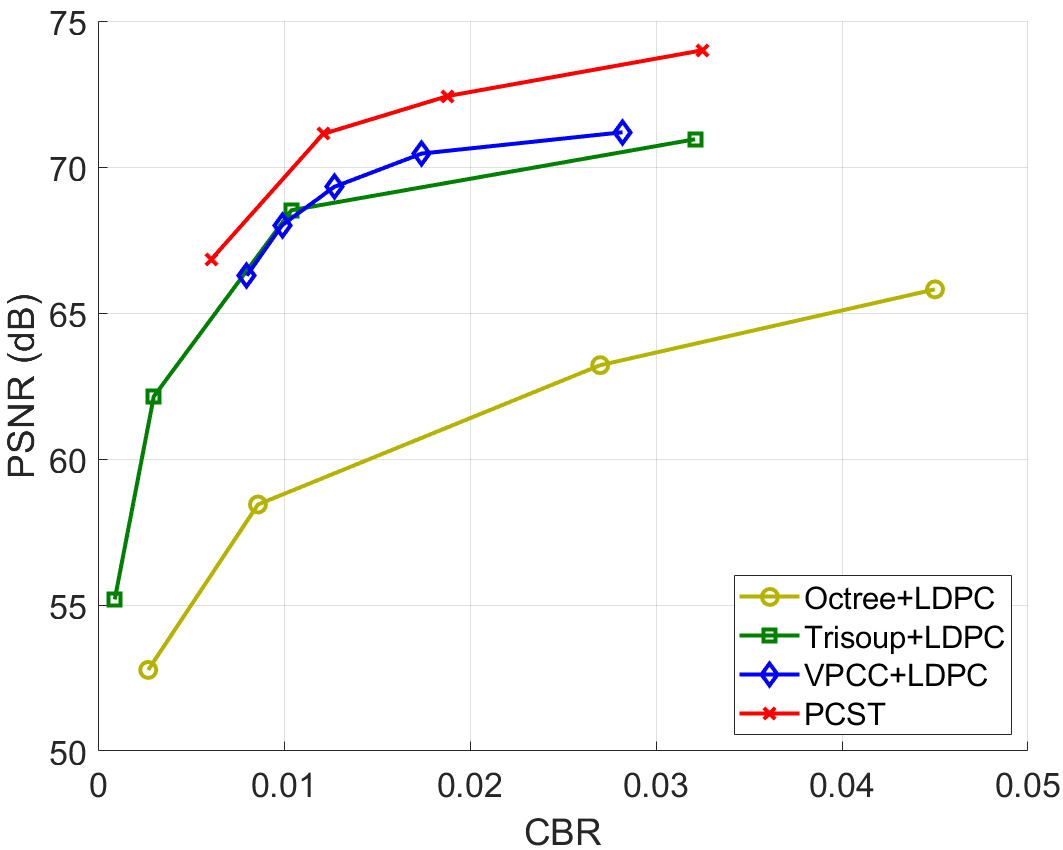}
    }
    \hfill
    \subfigure[]{
        \includegraphics[width=0.235\textwidth]{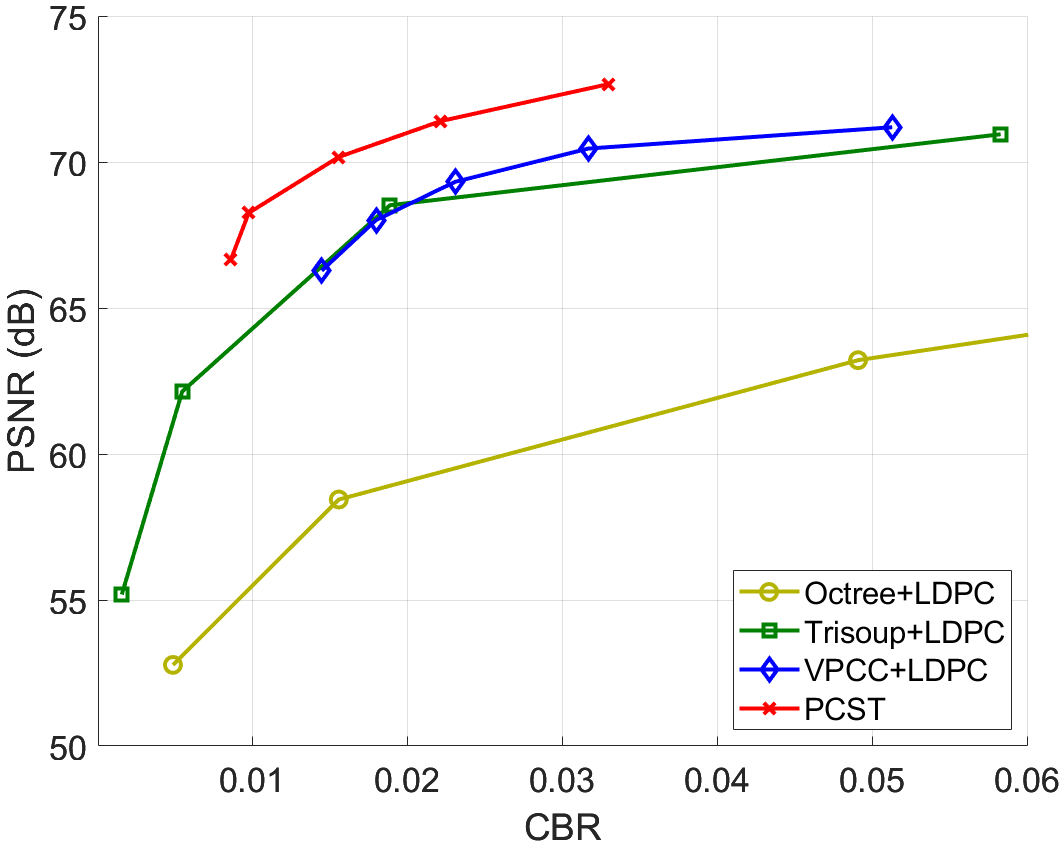}
    }
    \hfill
    \subfigure[]{
        \includegraphics[width=0.235\textwidth]{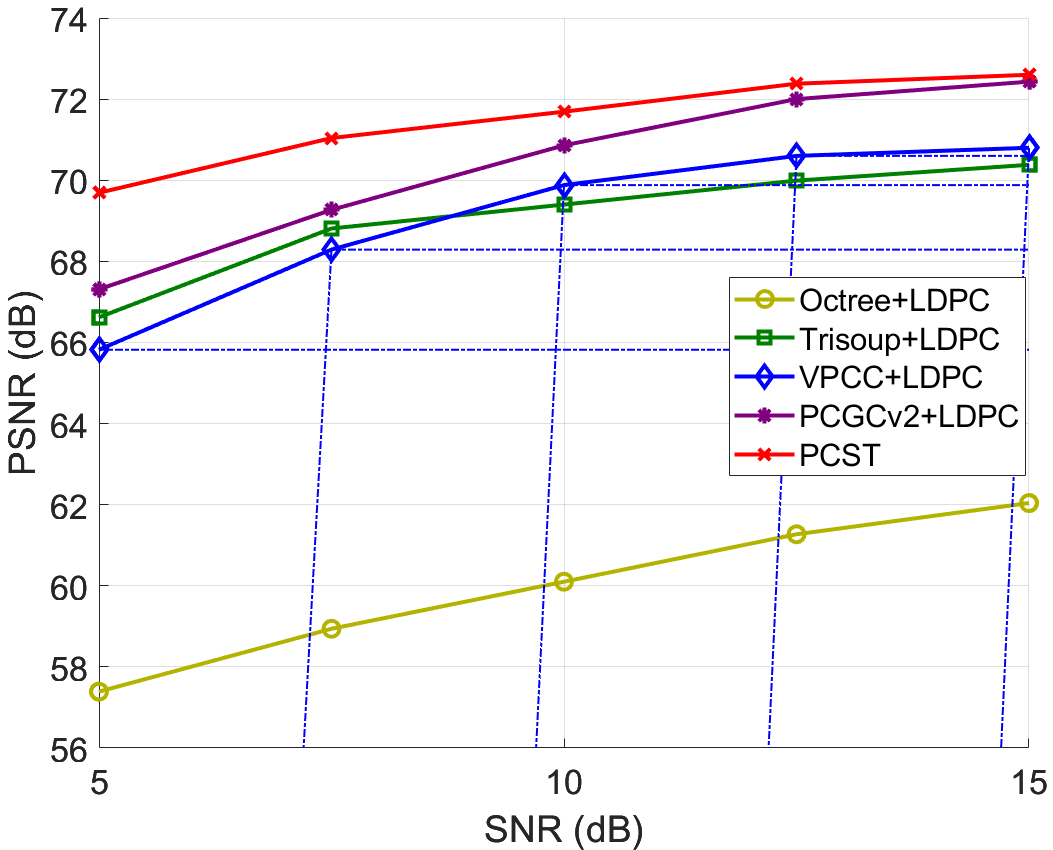}
    }
    \hfill
    \subfigure[]{
        \includegraphics[width=0.235\textwidth]{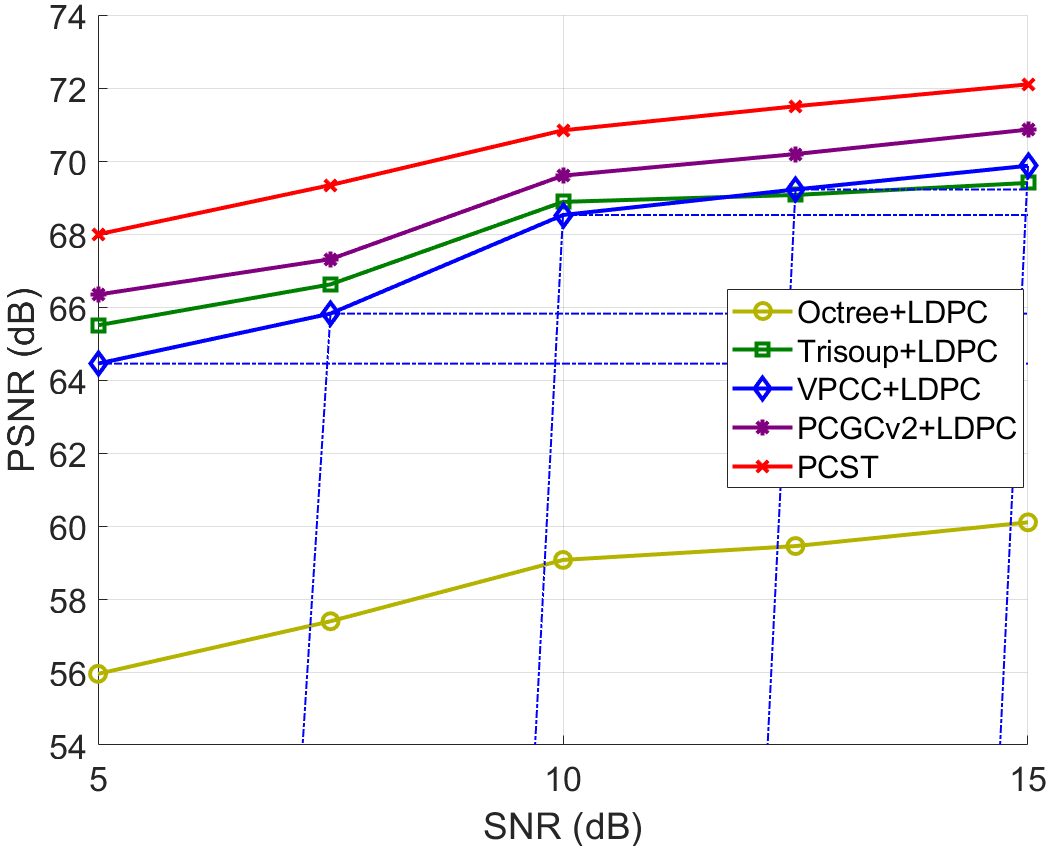}
    }
    \vspace{-3mm}
    \caption{\small PCST performance vs. CBR on D1-PSNR for (a) AWGN channel and (b) Rayleigh channel at SNR = 10 dB; PCST performance vs. SNR under fixed CBR for (c) AWGN channel (CBR=0.015) and (d) Rayleigh channel (CBR=0.020).}
    \label{fig}
       \vspace{-5mm}
\end{figure*}

\begin{table*}[t]
\small
\centering
\caption{\small Comparison of PCST performance in cross-data eval. Best in red, second best in orange, and third best in yellow.}
\label{tab:comparison}
\resizebox{1.99\columnwidth}{!}{
\begin{tabular}{c|cc|cc|cc|cc|cc|cc}
\hline
Point cloud & \multicolumn{4}{c|}{Soldier} & \multicolumn{4}{c|}{Loot} & \multicolumn{4}{c}{Red and black} \\ \hline 
Channel & \multicolumn{2}{c|}{AWGN} & \multicolumn{2}{c|}{Rayleigh} & \multicolumn{2}{c}{AWGN}  & \multicolumn{2}{c|}{Rayleigh} & \multicolumn{2}{c|}{AWGN} & \multicolumn{2}{c}{Rayleigh} \\ \hline 
Metrics & D1↑ & D2↑ & D1↑ & D2↑ & D1↑ & D2↑ & D1↑ & D2↑ & D1↑ & D2↑ & D1↑ & D2↑\\ \hline
GPCC (octree) + LDPC & 60.02 & 64.29 & 59.00 & 62.94 & 60.48 & 64.89 & 59.40 & 63.47 & 59.84 & 64.12 & 58.95 & 62.94 \\ 
GPCC (trisoup) + LDPC & \cellcolor{orange!25}69.46 & \cellcolor{orange!25}71.69 & \cellcolor{orange!25}68.90 & \cellcolor{orange!25}70.94 & \cellcolor{yellow!25}69.73 & \cellcolor{yellow!25}72.10 & \cellcolor{yellow!25}69.30 & \cellcolor{yellow!25}65.86 & \cellcolor{orange!25}68.96 & \cellcolor{orange!25}71.06 & \cellcolor{orange!25}68.35 & \cellcolor{orange!25}70.30 \\ 
VPCC + LDPC & \cellcolor{yellow!25}69.15 & \cellcolor{yellow!25}71.61 & \cellcolor{yellow!25}67.37 & \cellcolor{yellow!25}69.36 & \cellcolor{orange!25}70.42 & \cellcolor{orange!25}72.96 & \cellcolor{orange!25}69.41 & \cellcolor{orange!25}71.82 & \cellcolor{yellow!25}68.42 & \cellcolor{yellow!25}70.67 & \cellcolor{yellow!25}66.49 & \cellcolor{yellow!25}68.44 \\ 
PCST & \cellcolor{red!35}71.42 & \cellcolor{red!35}74.82 & \cellcolor{red!35}70.72 & \cellcolor{red!35}74.06 & \cellcolor{red!35}72.51 & \cellcolor{red!35}76.09 & \cellcolor{red!35}71.25 & \cellcolor{red!35}74.61 & \cellcolor{red!35}70.62 & \cellcolor{red!35}74.41 & \cellcolor{red!35}69.92 & \cellcolor{red!35}73.63 \\ 
\hline
\end{tabular}}
\vspace{-3mm}
\end{table*}


\subsection{Optimization goal} 
\small Throughout the optimization process, we need to trade off between channel bandwidth cost and the quality of the reconstructed point cloud. Thus, we define the rate-distortion loss as
\begin{equation}
    J_{\rm loss}=K +\lambda D.
\end{equation}
$K$ is the total channel bandwidth cost for latent features and $D$ is the PCG distortion. In PCG reconstruction, the generation of each voxel point is determined by predicting its occupancy. The probability that a voxel is occupied is computed using sparse convolution that leverages latent features from a lower scale. As this probability ranges between 0 and 1, the sigmoid function is employed for activation. The occupancy threshold can be adjusted to regulate the density of the reconstructed PCG. During the upscaling phase, we rank the occupancy probabilities and select the top-$k$ voxels, which are considered the most likely to be occupied. This process is often referred to as top-$k$ generation. Experiments have found that setting $k$ to the number of points in the ground truth label at each scale can achieve almost the lowest point-to-point distortion. The distortion $D$ is thus modeled as the Binary Cross-Entropy (BCE) loss, which is commonly used for classification task as
\begin{equation}
    D=\frac{1}{N}\sum_i -(x_i \log (p_i)+(1-x_i)\log(1-p_i)),
\end{equation}
where $x_i$ is the ground truth and $p_i$ denotes the predicted probability. By adjusting $\lambda$, we can train PCST models capable of producing transmitted symbols with varying code lengths.

\subsection{PCST overhead}
In addition to transmitting symbols $S$, PCST requires extra side information for PCG reconstruction. All side information, including the coordinates of latent features, the encoded length list, and the number of points for top-$K$ generation, is transmitted. The third component is omitted from the discussion as it involves only three integers per point cloud transmission. The coordinates of the latent features are transmitted to establish the PCG's basic contour, while the variable lengths of symbols require $k$-length information to map them to the corresponding FC layer. Fortunately, these side information minimally impacts channel bandwidth. In our experiments with AWGN channel at an SNR of 10 dB, we found that GPCC-encoded coordinates constitute only 0.003 CBR, and entropy-encoded $k$-length information accounts for just 0.00017 CBR.


\section{Experiment}

In this section, we assess PCST under various channel conditions and compare its reconstruction quality with conventional SSCC schemes, GPCC+LDPC and VPCC+LDPC.

\subsection{Implementation details}
\;\;\;\;\;1) Model details: We use sparse convolution to build a three-layer progressive resampling network for extracting semantic features of point clouds. Additionally, we employ a switchable FC layer module as the JSCC codec, shown in Fig. \ref{fig:enter-label}. The FC layer has an input dimension of 8, and its output dimension is chosen from the closest value in \(klist\) according to entropy predictions. We determine \(klist\) by examining the entropy prediction range using the compressed data from \cite{PCGCv2}.

\par 2) Dataset: We use the ShapeNet dataset \cite{shape} to train PCST, comprising approximately 51,300 CAD surface models. Before training, these models undergo dense sampling to create point clouds. Subsequently, they are randomly rotated and quantized to 7-bit precision for each dimension. The quantity of points in each point cloud is varied without restrictions. We randomly divide the dataset into two portions, 80\% for training and the remaining 20\% for validation. To assess the PCST model's performance in comparison to other SSCC methods, we use point cloud frames from four distinct sequences in 8i \cite{8i}, Soldier, Loot, and Red and black. 
\par 3) Benchmarks: We combine popular techniques for point cloud compression and channel coding to establish three SSCC benchmarks: VPCC+LDPC, GPCC (octree)+LDPC, and GPCC (trisoup)+LDPC. The quantization parameter and other key parameter settings follow the default configurations of the MPEG TMC framework \cite{tmc}, while the LDPC coding and modulation adhere to the adaptive modulation and coding mechanism \cite{glob, amc}.

\subsection{Performance comparison}
To compare the performance of PCST with other benchmarks under various conditions, we employ D1-PSNR (point-to-point) and D2-PSNR (point-to-plane) to measure the visual quality of the reconstructed PCG. It is important to note that due to the stochastic nature of the communication channel, we averaged the results over multiple tests to obtain the final results. We begin by illustrating the reconstruction qualities of longdress PCG in Fig. \ref{vis}, which shows that PCST efficiently reconstruct PCG with higher qualities at even lower CBRs. To gain further insight, we train PCST using varying \(\lambda\) values to produce point clouds with different CBRs, and conduct tests in both AWGN and Rayleigh channels. Fig. \ref{fig} (a) illustrates the reconstruction quality of PCST versus CBR in AWGN channel. From the figure, it is apparent that PCST outperforms the benchmarks in both high and low CBR conditions. Under the same reconstruction quality conditions, such as achieving a D1-PSNR of 70dB, PCST can reduce bandwidth usage by more than 50\% compared to other SSCC schemes. This means that for the same quality of point cloud reconstruction, PCST requires significantly less data transmission, leading to more efficient bandwidth usage. In Fig. \ref{fig} (b), the performance of all methods shows a decline due to the multiplicative factor \(h\) caused by the Rayleigh channel, leading to increased distortions. Nevertheless, PCST still exceeds the performance of all benchmarks, showcasing 2-10dB gains in reconstruction quality at CBR=0.035. Fig. \ref{fig} (c) (d) illustrate the performance of PCST under different SNRs with constrained CBR. It is noted that PCST displays more consistent performance changes, as opposed to SSCC which experiences the cliff effect. This is mainly because the latent features are resistant to noise, enabling it to reconstruct an acceptable point cloud structure even when certain features are missing. Consistently, PCST surpasses other SSCC methods with improvements ranging from 0.4 to 4 dB across various SNRs. Due to space constraints, a detailed analysis is not included here. To showcase the generalization ability of PCST, cross-data validation results for all methods are presented in Table 1. The table reveals that PCST consistently achieves higher reconstruction quality across different scenarios.

\section{Conclusion}
\small We introduce PCST, the first semantic transmission framework designed for point cloud transmission over wireless networks. PCST extracts the semantic characteristics of the point cloud and transmits them with joint source channel coding. Specifically, our approach utilizes a progressive resampling framework along with a deep JSCC network capable of adaptable bandwidth allocation to ensure high-quality PCG reconstruction. Experimental results demonstrate that our model outperforms traditional SSCC schemes in both AWGN and Rayleigh channels. Notably, the PCST reconstructs high-quality point clouds with significantly reduced bandwidth usage, making it a promising solution for metaverse communication systems.

\small 
\bibliographystyle{IEEEbib}
\bibliography{refs}

\end{document}